\documentstyle[preprint,prc,eqsecnum,aps,epsf]{revtex}

\tightenlines   

\begin{document}
\draft

\title{Shadowing and Antishadowing Effects in a Model for the n+d
Total Cross Section}

\author{ Ch.~Elster, W.~Schadow} 
\address{Institute of Nuclear and
Particle Physics, and Department of Physics, \\ Ohio University,
Athens, OH 45701}
    
\author{ H. Kamada\footnote{Present address:
Institut f\"ur Kernphysik, Fachbereich 5, Technische Hochschule
Darmstadt, D-64289 Darmstadt, Germany} and W. Gl\"ockle}
\address{ Institute for Theoretical Physics II, Ruhr-University Bochum,
D-44780 Bochum, Germany.}

\vspace{10mm}

\date{\today}

\maketitle

\begin{abstract}
Based on the multiple scattering series incorporated in the Faddeev
scheme the high energy limit of the total n+d cross section is
evaluated in a nonrelativistic model system where spins are neglected.
In contrast to the naive expectation that the total n+d scattering
cross section is the sum of two NN cross sections we find two
additional effects resulting from rescattering processes. These
additional terms have different signs (shadowing and antishadowing)
and a different behavior as function of the energy.  Our derivation of
these results which are already known from Glauber theory is based on
the analytical evaluation of elastic transition amplitudes in the high
energy limit.  It does not depend on the diffraction type assumptions
connected with Glauber theory.  In this model of spinless Yukawa type
forces (with no absorption) the total n+d cross section does not
approach twice the NN total cross section in the high energy limit
but rather approaches the
total NN cross section multiplied by a number larger than two.
Therefore, the enhancement effect resulting from
rescattering is larger than the shadowing effect, which decreases
faster with energy.
\end{abstract}

\vspace{10mm}


\pagebreak

 
\narrowtext
 

\section{Introduction}

Recently it became possible to calculate the total cross section for
neutron-deuteron (n+d) scattering with high precision in the energy
regime up to 300~MeV projectile energy by solving the nonrelativistic
3N Faddeev equations based on modern NN forces \cite{ndtotl}.
Compared to the most naive picture, which in the high energy limit
would equate the total n+d cross section with the sum of the total
cross sections for neutron-proton (np) and neutron-neutron (nn)
scattering, the rigorously calculated result up to 300 MeV is
smaller. Obviously one can expect some shadowing effect in the
reaction, which would explain this result.  On the other hand,
rescattering of the nucleons upon each other might a priori also
enhance the total n+d cross section over the sum of the individual two
nucleon cross sections, especially if the forces are attractive.  In
the context of Glauber theory \cite{glauber,joachain,henley,abers}
both features, enhancement and weakening, are present. In a model of
spinless particles, which is based on the Faddeev formulation, we want
to study the high energy limit of the total n+d cross section and
evaluate the leading terms analytically.  This approach differs from
the one used in the Glauber formulation.  Based on the multiple
scattering series incorporated in the Faddeev framework, we calculate
the first and second order terms of the series when taken in the high
energy limit and show the contributions of the two terms to the total
n+d cross section.  Starting from a multiple scattering series implies
that this is an ordering according to powers in the NN $t$-matrix.
Though we restrict ourselves to a nonrelativistic framework, we
nevertheless consider the results as being instructive.  The
analytical steps leading to the high energy limit, which only involve
ordinary analysis, are carried out in well defined integrals. There
are no a priori assumptions about the scattering process involved,
like e.g., diffraction type approximations.  We also take the identity
of the particles explicitly into account.  In this paper the
complications due to the inclusion of the spin and isospin degrees of
freedom are avoided so that the basic mechanism can be seen more
clearly. Taking spin and isospin degrees of freedom into account will
modify the results due to the interferences of spin and isospin
dependent terms, as will be shown in a forthcoming
article. Specifically because of those spin and isospin interference
effects, which are quite involved, we want to present this more
transparent case with three bosons separately.

In Section II we describe the Faddeev framework, its multiple
scattering expansion, and the leading order terms in the NN $t$-matrix
for obtaining the total n+d cross section. The high energy limit of
the corresponding expressions is carried out analytically in Section
III. To illustrate the behavior of the leading order terms in the high
energy limit numerical examples are given in Section IV for a
superposition of Yukawa interactions. We conclude with Section V.

\section{Leading Multiple Scattering Terms for the Total n+d Cross Section}

We consider three identical spinless bosons which interact via
two-body forces. In our usual manner \cite{Wbook} to exploit the
Faddeev scheme the operator for elastic scattering of a nucleon from a
bound nucleon pair is given by~\cite{physrep}
\begin{equation}
U = P G_0^{-1} +PT,  \label{eq:2.1}
\end{equation}
where the three-body operator $T$ obeys the Faddeev type integral
equation
\begin{equation}
T |\Phi \rangle = tP|\Phi \rangle + tPG_0T|\Phi \rangle. \label{eq:2.2}
\end{equation}
The channel state, which is composed of a two nucleon bound state
(deuteron) and a momentum eigenstate of the projectile nucleon, is
denoted by $\Phi$. Furthermore, $t$ is the two nucleon transition
operator, $G_0$ the free three nucleon propagator, and $P$ the
permutation operator, which is the sum of a cyclic and an anticyclic
permutation of the three particles. The elastic forward scattering
amplitude is given by the matrix element $\langle \Phi|U|\Phi
\rangle$.  When iterating Eq.~(\ref{eq:2.2}) and inserting the result
into Eq.~(\ref{eq:2.1}), we obtain the multiple scattering series for
the elastic forward scattering amplitude
\begin{eqnarray}
\langle \Phi |U|\Phi \rangle &=& \langle \Phi |PG_0^{-1}|\Phi \rangle
+ \langle \Phi | PtP |\Phi \rangle + \langle \Phi |PtPG_0 tP|\Phi
\rangle \nonumber \\ & & + \, \langle \Phi |PtP G_0 tP G_0 tP |\Phi
\rangle + \cdots , \label{eq:2.3}
\end{eqnarray}
which is an expansion in orders of the NN $t$-operator.  Using the
optical theorem one obtains for the total cross section for
nucleon-deuteron (n+d) scattering \cite{physrep}
\begin{equation}
\sigma^{nd}_{tot} = -(2\pi)^3 \frac{4m}{3q_0} Im \langle \Phi |U| \Phi
 \rangle.  \label{eq:2.4}
\end{equation}
Here $q_0$ is the asymptotic momentum of the projectile nucleon
relative to the bound two-body subsystem. From Eq.~(2.3) follows
\begin{eqnarray}
2 i \; Im \langle \Phi |U|\Phi \rangle &\equiv &
\langle \Phi |U|\Phi \rangle - \langle \Phi |U|\Phi \rangle^{\star}
 \label{eq:2.5} \nonumber \\
&=& \langle \Phi |  P(t-t^{\dagger})P + PtP G_0 tP - 
P t^{\dagger} G_0^{\star}Pt^{\dagger} P +PtP G_0 tP G_0 tP 
 \nonumber \\
& & - \, P t^{\dagger} G_0^{\star} Pt^{\dagger}G_0^{\star} Pt^{\dagger}P 
| \Phi \rangle + \cdots \;\; . 
\end{eqnarray}
Since the first term in Eq.~(2.3) is real, it does not contribute to
the total cross section.

For the analytical extraction of the high energy limit it is useful to
rewrite Eq.~(2.5) in the following form
\begin{eqnarray}
\langle \Phi |U|\Phi \rangle &-& \langle \Phi |U|\Phi \rangle^{\star}
 \nonumber \\ &=& \langle \Phi | P(t-t^{\dagger})P|\Phi \rangle +
 \langle \Phi | P(t-t^{\dagger})P G_0 tP |\Phi \rangle \nonumber \\ &
 & - \, \langle \Phi |P(t-t^{\dagger})P G_0 tP |\Phi \rangle^{\star} +
 \langle \Phi | Pt^{\dagger} P(G_0-G_0^{\star})tP |\Phi \rangle
 \nonumber \\ & & + \, \langle \Phi |P(t-t^{\dagger})P G_0 tP G_0
 tP|\Phi \rangle \nonumber \\ & & + \, \langle \Phi |P t^{\dagger}
 P(G_0-G_0^{\star})tPG_0 tP|\Phi \rangle \nonumber \\ & & + \, \langle
 \Phi |P t^{\dagger} PG_0^{\star}(t-t^{\dagger})P G_0 tP |\Phi \rangle
 \nonumber \\ & & + \, \langle \Phi |P t^{\dagger}
 PG_0^{\star}t^{\dagger} P(G_0-G_0^{\star}) tP|\Phi \rangle \nonumber
 \\ & & + \, \langle \Phi |P t^{\dagger} PG_0^{\star}t^{\dagger}
 PG_0^{\star} (t-t^{\dagger})P |\Phi \rangle + \cdots \;\;
 . \label{eq:2.6}
\end{eqnarray}
The next step is to explicitly evaluate the permutations given by
$P\equiv P_{12}P_{23}+P_{13}P_{23}$. This specific choice of the
permutation operator $P$ corresponds to the choice of particles 2 and
3 forming the two-body bound state and 1 being the projectile, i.e.,
$t\equiv t_{23}$ and
\begin{equation}
 |\Phi \rangle \equiv |\Phi \rangle_1 \equiv |\varphi_d \rangle_{23}
|{\bf q_0}\rangle_1 . \label{eq:2.7}
\end{equation}
The subscripts denote which particles occupy the states.  This
specific choice leads to
\begin{equation}
|\Phi \rangle_2 \equiv P_{12}P_{23} |\Phi \rangle_1 = |\varphi_d
\rangle_{13} |{\bf q_0}\rangle_2 \label{eq:2.8}
\end{equation}
and
\begin{equation}
|\Phi \rangle_3 \equiv P_{13}P_{23} |\Phi \rangle_1 = |\varphi_d
\rangle_{12} |{\bf q_0}\rangle_3. \label{eq:2.9}
\end{equation}
Taking advantage of the symmetry property of the subsystem, $
 P_{23}|\Phi\rangle_1 =|\Phi \rangle_1$ and $P_{23} t_{23} = t_{23}
 P_{23}$, one obtains after some algebra
\begin{equation}
\langle \Phi |P (t-t^{\dagger})P |\Phi \rangle =
2 \:\: _2\langle \Phi |(t-t^{\dagger})(1+P_{23})|\Phi \rangle_2
 \label{eq:2.10a}
\end{equation}
and
\begin{equation}
\langle \Phi |P tPG_0 tP|\Phi \rangle 
= 2 \:\: _2\langle \Phi |tG_0 t_2\left( |\Phi \rangle_1 + |\Phi \rangle_3
\right) +2 \:\: _2\langle \Phi |tG_0 t_3 \left( |\Phi \rangle_1 + |\Phi 
 \rangle_2 \right).   \label{eq:2.10b}
\end{equation}
A similar evaluation could be done for the terms third order in $t$
given in Eq.~(\ref{eq:2.5}). For our present considerations of the n+d
total cross section in the high energy limit we restrict ourselves to
the terms given in Eqs.~(\ref{eq:2.10a}) and (\ref{eq:2.10b}).

\section{The High Energy Limit}

First we need to consider the exact momentum space representations of
 Eqs.~(\ref{eq:2.10a}) and (\ref{eq:2.10b}). Equivalent decompositions
 of the unity operator are given by
\begin{equation}
{\bf 1} = \int d^3p\; d^3q \; |{\bf p}{\bf q}\rangle_i \; _i\langle
 {\bf p}{\bf q}| \;\;\;,\;\;\; i=1,2, \, \text{or} \;
 3. \label{eq:2.11}
\end{equation}
Here $|{\bf p}{\bf q}\rangle_i \equiv | {\bf p}\rangle_i |{\bf
q}\rangle_i$ and ${\bf p}$ and ${\bf q}$ are the standard three types
of Jacobi momenta for three particles \cite{Wbook}. The index $i$
denotes the singled out nucleon.  Inserting the unity operator into
the first term in Eq.~(\ref{eq:2.10a}) leads to
\begin{eqnarray}
_2\langle \Phi | t-t^{\dagger}|\Phi \rangle_2 &=&\int d^3p \; d^3q
\int d^3p' d^3q' \int d^3p'' d^3q''\int d^3p''' d^3q''' \;\; _2\langle
\Phi|{\bf p}{\bf q}\rangle_2 \; _2\langle {\bf p}{\bf q}| {\bf p'}{\bf
q'}\rangle_1 \nonumber \\ & & \times \, _1\langle {\bf p'}{\bf
q'}|t-t^{\dagger}|{\bf p''}{\bf q''}\rangle_1 \; _1\langle {\bf
p''}{\bf q''}|{\bf p'''}{\bf q'''}\rangle_2 \; _2\langle {\bf
p'''}{\bf q'''}| \Phi \rangle_2 . \label{eq:2.12}
\end{eqnarray}
Further, one has
\begin{equation}
_i\langle {\bf p}{\bf q}|\Phi \rangle_i = \delta^3 ({\bf q} - {\bf
q_0}) \, \langle {\bf p}| \varphi_d \rangle . \label{eq:2.13}
\end{equation}
The standard relations among the different sets of Jacobi momenta give
\begin{equation}
_1\langle {\bf p}{\bf q}|{\bf p'}{\bf q'}\rangle_2 = \delta^3 ({\bf p}
- {\textstyle \frac {1}{2}}{\bf q} -{\bf q'}) \, \delta^3 ({\bf p'} +{\bf
q} + {\textstyle \frac {1}{2}} {\bf q'}) . \label{eq:2.14}
\end{equation}
In addition one has
\begin{equation}
_1\langle {\bf p}{\bf q}|t(E)|{\bf p'}{\bf q'}\rangle_1 = \langle {\bf
p}| t(E-{\textstyle \frac{3}{4m}} q^2)|{\bf p'}\rangle \, \delta^3({\bf
q}- {\bf q'}). \label{eq:2.15}
\end{equation}
Employing all the above given relations, it is straightforward to
arrive at the following expression for Eq.~(3.2)
\begin{eqnarray}
_2\langle \Phi | t-t^{\dagger}|\Phi \rangle_2 &=& \int d^3q \; \langle
\varphi_d| -{\bf q} -{\textstyle \frac{1}{2}} {\bf q_0} \rangle \, \langle
{\textstyle \frac{1}{2}} {\bf q}+{\bf
q_0}|t-t^{\dagger}|{\textstyle \frac{1}{2}} {\bf q} +{\bf q_0} \rangle \,
\langle -{\bf q} - {\textstyle \frac{1}{2}} {\bf q_0}| \varphi_d
\rangle  \nonumber \\ &=& \int d^3k \; \langle \varphi_d| -{\bf
k}\rangle \, \langle -{\bf k}| \varphi_d \rangle \, \langle
{\textstyle \frac{3}{4}} {\bf q_0} + {\textstyle \frac{1}{2}} {\bf k}|
(t-t^{\dagger}) (\varepsilon)| {\textstyle \frac{3}{4}} {\bf q_0} +
{\textstyle \frac{1}{2}} {\bf k} \rangle .  \label{eq:2.16}
\end{eqnarray}
The total energy is $E=\epsilon_d +\frac{3}{4m}q_0^2$, where
$\epsilon_d$ is the deuteron binding energy. When expressed in terms
of ${\bf k}={\bf q}+\frac{1}{2} {\bf q_0}$ one obtains for the energy
argument $\varepsilon$ of the $t$-matrices
\begin{eqnarray}
\varepsilon &\equiv& E-{\textstyle \frac{3}{4m}}q^2 = \epsilon_d
+{\textstyle \frac{3}{4m}}q_0^2 -{\textstyle \frac{3}{4m}}({\bf
k}-{\textstyle \frac {1}{2}}{\bf q_0})^2 \nonumber \\ &=& \epsilon_d +
{\textstyle \frac{1}{m}} \left ( {\textstyle \frac{3}{4}}{\bf
q_0}+{\textstyle \frac {1}{2}}{\bf k} \right )^2
-{\textstyle \frac{1}{m}} k^2. \label{eq:2.17}
\end{eqnarray}
If the projectile momentum $q_0$ is sufficiently large in comparison
to the dominant deuteron momenta contributing to the integral in
Eq.~(3.6) and thus $\varepsilon \approx \frac{1}{m} (\frac{3}{4}{\bf
q_0} +\frac {1}{2}{\bf k})^2$, one encounters on shell NN forward
scattering amplitudes under the integral. The permutation operator
$P_{23}$ in Eq.~(\ref{eq:2.10a}) leads to the necessary symmetrization
of the $t$-matrix elements
\begin{equation}
\langle {\bf q} \ | (t-t^{\dagger}) ( \varepsilon =
{\textstyle \frac{1}{m}}q^2) | {\bf q} \rangle_s, \label{eq:2.17a}
\end{equation}
where 
\begin{equation}
|{\bf q} \rangle_s = | {\bf q} \rangle + |-{\bf q} \rangle,
\label{eq:2.17b}
\end{equation}
and ${\bf q} = {3 \over 4} {\bf q}_0 + {1 \over 2} {\bf k}$. Using the
Lippmann-Schwinger equation for $t$, with the two-body force $V$ as
driving term, and ${\bf q}\equiv |q| {\bf {\hat q}}$ one has
\begin{eqnarray}
\langle {\bf q} \ | (t-t^{\dagger}) ( \varepsilon = {\textstyle
\frac{1}{m}}q^2) | {\bf q} \rangle_s & = & - 2 \pi i \, {m \over 2} \,
q \, \int d {\hat q}' \; \langle {\bf q} | V | q {\bf {\hat q}}'
\rangle^{(+)} \ ^{(+)}\langle q {\bf {\hat q}}' | V | {\bf q }
\rangle_s \nonumber \\ &=& - 2 \pi i \, {m \over 2} \, q \, \int d
{\hat q}' \; \langle {\bf q} | t | q {\bf {\hat q}}' \rangle \;
_s\langle {\bf q} | t | q {\bf {\hat q}}' \rangle^* \nonumber \\ &=& -
\pi i \, {m \over 2} \, q \, \int d {\hat q}' \; \left | \langle {\bf
q} | t | q {\bf {\hat q}}' \rangle_s \right |^2 .  \label{eq:2.17c}
\end{eqnarray}  
This expression is directly related to the two-body total cross
section
\begin{equation}
\sigma_{tot}^{NN} = \left( \frac{m}{2} \right)^2 (2\pi)^4 \int d {\hat
q}' \; \left | \langle {\bf q} | t | q {\bf {\hat q}}' \rangle_s { 1
\over \sqrt{2}} \right |^2 . \label{eq:2.17d}
\end{equation}
We adopted here the usual convention for the total cross section for
 two identical particles \cite{goldberger} and obtain
\begin{equation}
\langle {\bf q} | (t-t)^{\dagger}) (\varepsilon =
{{\textstyle \frac{1}{m}}}q^2) | {\bf q} \rangle_s = -i q \, {2 \over m}
{1 \over (2 \pi)^3} \, \sigma_{tot}^{NN}. \label{eq:2.17e}
\end{equation}
As a consequence Eq.~(\ref{eq:2.16}) supplemented by the
symmetrization as given in Eq.~(\ref{eq:2.10a}) and approximated in
the energy argument of the two-body $t$-matrix can be rewritten as
\begin{eqnarray}
\lefteqn{_2\langle \Phi|(t-t^{\dagger})(1+P_{23})|\Phi \rangle_2 }
\label{eq:2.18} \hspace{2cm} \\
\nonumber & = & - \frac{2i}{m} \frac{1}{(2\pi)^3} \int d^3k \; \langle
\varphi_d| -{\bf k}\rangle \, \langle -{\bf k}|\varphi_d \rangle \, |
{\textstyle \frac{3}{4}}{\bf q_0} + {\textstyle \frac{1}{2}}{\bf k}|
\, \sigma_{tot}^{NN} \left ({\textstyle \frac{1}{m}}({\textstyle
\frac{3}{4}}{\bf q_0}+ {\textstyle \frac{1}{2}}{\bf k})^2 \right ).
\nonumber
\end{eqnarray}
If the projectile momentum $q_0$ is sufficiently large compared to the
typical deuteron momenta the function $|\frac{3}{4}{\bf q_0} +\frac
{1}{2}{\bf k}| \sigma_{tot}^{NN} (\frac{1}{m}(\frac{3}{4}{\bf
q_0}+\frac {1}{2}{\bf k})^2)$ is expected to vary slowly over the
range of the ${\bf k}$-values contributing to the integral in
Eq.~(3.13). Using the normalization of the two-body bound state,
expanding the function at ${\bf k}=0$ and knowing that the
contribution of the first derivative vanishes, one obtains in the
limit for large ${\bf q_0}$
\begin{eqnarray}
_2\langle \Phi |(t-t^{\dagger})(1+P_{23})|\Phi \rangle_2 & 
\stackrel{q_0 \gg k}{\longrightarrow}& 
- \frac{2 i}{m} \frac{1}{(2\pi)^3} \frac{3}{4} \, q_0 \,\sigma_{tot}^{NN}
\left ({\textstyle \frac{1}{m}(\frac{3}{4}} q_0)^2 \right ) \int d^3k
\; \langle \varphi_d|-{\bf k}\rangle \, \langle -{\bf k}|\varphi_d \rangle
 \nonumber \\
&=& - \frac{2 i}{m} \frac{1}{(2\pi)^3} \frac{3}{4} \, q_0 \, \sigma_{tot}^{NN}
\left ({\textstyle \frac{1}{m}(\frac{3}{4}} q_0)^2 \right ).
\label{eq:2.19}
\end{eqnarray}

The arguments leading to Eq.~(3.14) are similar to those used in
arriving at the method of optimum factorization successfully applied
in intermediate energy pion-nucleus and nucleon-nucleus scattering
\cite{optf,wolfe}. The n+d total cross section, Eq.~(\ref{eq:2.4}),
thus gives in the first order term in $t$ of Eq.~(2.6) in the high
energy limit
\begin{equation}
\sigma^{nd}_{tot}|_{1st\:{\rm order}}
 \rightarrow  2 \, \sigma^{NN}_{tot}. \label{eq:2.20}
\end{equation}

This result corresponds to the naive expectation that at high energies
the projectile nucleon sees the individual nucleons inside the
deuteron as if they were independent particles. It should be pointed out
that due to the optical theorem $\sigma^{NN}_{tot}$ is $O(t^2)$, though
the expression of Eq.~(\ref{eq:2.20}) has been derived from the 
terms linear in $t$ in Eq.~(\ref{eq:2.6}).

Let us now consider the contributions to the total n+d cross section
second order in $t$ as given in Eq.~(\ref{eq:2.5}). A straightforward
but somewhat tedious algebra using Eq.~(2.11) leads to the exact form
\begin{eqnarray}
\lefteqn{\langle \Phi | P\tau PG_0 tP | \Phi \rangle} \nonumber \\ &=&
2 \int d^3q \int d^3q' \, \frac {\langle \varphi_d| {\bf q}\rangle
\langle {\bf q'}|\varphi_d \rangle} {-|\epsilon_d|
-\frac{1}{m}(q^2+q'^2+{\bf q}\cdot{\bf q'}) +\frac{3}{2m}({\bf q}+{\bf
q'})\cdot {\bf q_0} +i\varepsilon} \nonumber \\ & & \;\;\times \langle
{\textstyle \frac{3}{4}} {\bf q_0}+{\textstyle \frac{1}{2}}{\bf q}|
\tau (\epsilon_1)| {\textstyle \frac{3}{4}} {\bf q_0} - {\textstyle
\frac{1}{2}}{\bf q}-{\bf q'}\rangle \, \langle - {\textstyle
\frac{3}{4}}{\bf q_0}+{\bf q}+ {\textstyle \frac{1}{2}}{\bf q'}
|t(\epsilon_3)| {\textstyle \frac{3}{4}}{\bf q_0}+ {\textstyle
\frac{1}{2}}{\bf q'}\rangle_s \nonumber \\ & & + \, 2 \, \int d^3q
\int d^3q' \, \frac {\langle \varphi_d| {\bf q}\rangle \langle {\bf
q'}|\varphi_d \rangle} {-|\epsilon_d| -\frac{1}{m}(q^2+q'^2+{\bf
q}\cdot{\bf q'}) +\frac{3}{2m}({\bf q}+{\bf q'})\cdot {\bf q_0}
+i\varepsilon} \nonumber \\ & & \;\; \times \,\langle {\textstyle
\frac{3}{4}} {\bf q_0}+{\textstyle \frac{1}{2}}{\bf q}| \tau
(\epsilon_1)| -{\textstyle \frac{3}{4}} {\bf q_0} + {\textstyle
\frac{1}{2}}{\bf q} +{\bf q'}\rangle \, \langle {\textstyle
\frac{3}{4}}{\bf q_0}-{\bf q}- {\textstyle \frac{1}{2}}{\bf q'}
|t(\epsilon_2)|-{\textstyle \frac{3}{4}}{\bf q_0} - {\textstyle
\frac{1}{2}}{\bf q'}\rangle_s
\label{eq:2.21}
\end{eqnarray}
The quantity $\tau$ represents either $t-t^{\dagger}$ or
$t^{\dagger}$, and thus acts in the subsystem 1=(23). Again, the
subscript $s$ indicates the symmetrized state as given in
Eq.~(\ref{eq:2.17b}).  The energy arguments in the $t$-matrices are
\begin{equation}
\epsilon_1 = \epsilon_d + {\textstyle {1 \over m}} \left( {\textstyle {3
\over 4}} {\bf q}_0 \right) ^2 + {\textstyle {3 \over 4m}} {\bf q}
\cdot {\bf q}_0 - {\textstyle {3 \over 4m}} q^2
\end{equation}
and
\begin{equation}
\epsilon_3 = \epsilon_2 = \epsilon_d + {\textstyle  {1 \over m}} \left(
{\textstyle {3 \over 4}} {\bf q}_0 \right) ^2 + {\textstyle {3 \over
4m}} {\bf q}' \cdot {\bf q}_0 - {\textstyle {3 \over 4m}} q'^2
\end{equation}
It should be pointed out that the occurrence of the symmetrized state
incorporates the forward and backward scattering amplitudes.

For the limit $q_0 \rightarrow \infty$ one can again neglect the
variations of the $t$-matrices under the integrals and obtains in this
limit
\begin{equation}
\langle \Phi | P\tau PG_0 tP | \Phi \rangle 
\approx  2 \, \langle {\textstyle \frac{3}{4}} {\bf q_0} |\tau(
{\textstyle \frac{1}{m}}({\textstyle \frac{3}{4}} q_0)^2)|
{\textstyle \frac{3}{4}} {\bf q_0} \rangle_s \; \langle
{\textstyle \frac{3}{4}} {\bf q_0}
|t({\textstyle \frac{1}{m}}({\textstyle \frac{3}{4}}q_0)^2)|
{\textstyle \frac{3}{4}} {\bf q_0} \rangle_s \times I, \label{eq:2.22}
\end{equation}
where
\begin{equation}
I=\int d^3q \int d^3q' \frac {\langle \varphi_d| {\bf q}\rangle \,
\langle {\bf q'}|\varphi_d \rangle} {-|\epsilon_d|
-\frac{1}{m}(q^2+q'^2+{\bf q}\cdot{\bf q'}) +\frac{3}{2m}({\bf q}+{\bf
q'})\cdot {\bf q_0} +i\varepsilon} .
\label{eq:2.23}
\end{equation}

\noindent
The interesting insight will now arise from the quantity $I$, which we
still need to consider in the high energy limit. These considerations
are most transparent if one works in configuration space.  The
momentum space form will be given in Appendix A. Let
\begin{equation}
\langle {\bf q}|\varphi_d \rangle \equiv \frac{1}{(2\pi)^{3/2}} \int
 d^3r \; e^{i {\bf q}\cdot {\bf r}} \, \langle {\bf r}|\varphi_d \rangle
\label{eq:2.24}
\end{equation}
with
\begin{equation}
\langle {\bf r}|\varphi_d \rangle \equiv \frac{1}{\sqrt{4\pi}} \,
\varphi_d(r).
\label{eq:2.25}
\end{equation}
Then the quantity $I$ from Eq.~(\ref{eq:2.23}) becomes
\begin{eqnarray}
I &=& \frac{1}{(2\pi)^3} \frac{1}{4\pi} \int d^3q \int d^3q' \int d^3
  r \; \int d^3r'\; e^{i {\bf q}\cdot {\bf r}} \, e^{-i {\bf q'}\cdot
  {\bf r'}} \nonumber \\ & & \times \, \frac {\varphi_d(r) \,
  \varphi_d(r')} {-|\epsilon_d| -\frac{1}{m}(q^2+q'^2+{\bf q}\cdot{\bf
  q'}) +\frac{3}{2m}({\bf q}+{\bf q'})\cdot {\bf q_0} +i\varepsilon}.
\label{eq:2.26}
\end{eqnarray}

\noindent
Since the denominator depends on ${\bf q_0}$, the quantity whose
infinite limit we want to consider, it is natural to split the vector
${\bf q}$ (${\bf q}'$) in components parallel and perpendicular to
${\bf q_0}$, and consider ${\bf q_0}$ pointing into the
$z$-direction. For the vector ${\bf r}$ (${\bf r}'$) we define a
similar decomposition
\begin{eqnarray}
{\bf q} &\equiv& ({\bf q}_{\perp},q_z) \\ \nonumber
{\bf r} &\equiv& ({\bf r}_{\perp},z).\label{eq:2.27}
\end{eqnarray}
and obtain
\begin{eqnarray}
I &=& \frac{m}{(2\pi)^3} {1 \over 4 \pi}  \int d^2 q_{\perp} 
\int d^2 q'_{\perp}  
 \int d^3 r \; \varphi_d(r) \int d^3 r' \; \varphi_d(r') \;
 e^{i {\bf q}_{\perp}\cdot {\bf r}_{\perp}} \;
   e^{-i {\bf q'}_{\perp}\cdot {\bf r'}_{\perp}}  \\ \nonumber
 & & \times  \int d q_z \int d q_{z'} \;
  e^{i q_z z}  \; e^{-i q'_z z'}  \\ \nonumber
& &\times \, \frac{1}{ -m|\epsilon_d| - q_{\perp}^{2} - {q'}_{\perp}^2 -
{\bf q}_{\perp} \cdot {\bf q'}_{\perp} -q_z^2 -{q'}_z^2 -q_z q'_z
+\frac{3}{2} q_0 (q_z+q_z') +i\varepsilon}. \label{eq:2.28}
\end{eqnarray} 

\noindent
Further we substitute
\begin{eqnarray}
q_z +q'_z &=& s \\ \nonumber
{\textstyle \frac{1}{2}}(q_z -q'_z) &=& s' \label{eq:2.29}
\end{eqnarray}
so that the denominator takes the form
\begin{equation}
D= - {\textstyle \frac{3}{4}} (s^2 - 2q_0 s +
{\textstyle\frac{4}{3}}\alpha^2 -i\varepsilon),
\label{eq:2.31}
\end{equation}
with 
\begin{equation}
\alpha^2 = m|\epsilon_d| + q_{\perp}^2 +{q'}_{\perp}^2 + {\bf
q}_{\perp} \cdot {\bf q'}_{\perp} +{s'}^2 \;\; > 0
\label{eq:2.32}
\end{equation}

\noindent
For large $q_0$ one obtains
\begin{eqnarray}
D &\approx& - {\textstyle\frac{3}{4}} ( s - 2q_0 - i\varepsilon
 ) ( s - \frac{2}{3} \frac{\alpha^2}{q_0}
 +i\varepsilon ) \nonumber \\ &\approx& - {\textstyle\frac{3}{4}}
 ( s - 2q_0 -i\varepsilon ) ( s +i\varepsilon )
\label{eq:2.33}
\end{eqnarray}
and thus encounters two poles in the complex s-plane. The integrals
over $s$ and $s'$ can be carried out analytically, leading to
\begin{eqnarray}
\lefteqn{\int\limits^{\infty}_{-\infty}ds \int\limits^{\infty}_{-\infty} ds'
\; e^{i (z+z')s'} e^{\frac{i}{2}(z-z')s} \,  \frac{-4/3}
 {(s-2q_0-i\varepsilon) (s+i\varepsilon)}} \nonumber \hspace{2cm}\\ & =&
 - \frac{2i}{3q_0}(2\pi)^2 \delta(z+z') \left[ \theta(z) e^{2iq_0 z} +
 \theta(-z)\right].
\label{eq:2.34}
\end{eqnarray}
Inserting this into Eq.~(3.25) and carrying out the integrations
over ${\bf q_{\perp}}$ and ${\bf q_{\perp}'}$ yields
\begin{eqnarray}
I & \approx& m \, (2\pi)^3 \, \frac {-2i}{3q_0} \int d^2 r_{\perp}
\int d^2 {r'}_{\perp} \int\limits_{-\infty}^{\infty} dz
\int\limits_{-\infty}^{\infty} dz' \\ \nonumber & & \times \,
\varphi_d \left (\sqrt{r_{\perp}^2 +z^2} \right ) \,\varphi_d \left
(\sqrt{{r'}_{\perp}^2 +{z'}^2} \right ) \, \delta({\bf r}_{\perp}) \,
\delta({\bf r'}_{\perp}) \, \delta(z+z') \left[ \theta(z) e^{2i q_0 z}
+ \theta(-z) \right] \frac{1}{4\pi}. \label{eq:2.35}
\end{eqnarray}

\noindent
The $\delta$-functions under the integral show that contributions to
$I$ only arise if the two nucleons in the deuteron sit behind each
other with respect to the projectile momentum ${\bf q_0}=q_0 {\hat
z}$.  This coincides with our naive understanding of shadowing.  The
term containing $e^{2i q_0 z}$ falls off fastest in Eq.~(3.31) and
will be neglected. Thus, one arrives at
\begin{eqnarray}
I & \approx& -2 i m (2\pi)^3 \frac{1}{3q_0} \int\limits_{0}^{\infty} dz
\,  \varphi_d^2 (z) \frac{1}{4\pi}  \nonumber \\
&=& - \frac{2 i m}{3q_0} (2\pi)^3 \, \langle \varphi_d |\frac{1}{r^2}|
 \varphi_d \rangle  \frac{1}{4\pi} . \label{eq:2.36}
\end{eqnarray}
The corresponding algebraic steps performed in momentum space are
shown in Appendix A and yield the result
\begin{equation}
I \rightarrow -\frac{2i m}{3q_0} (2\pi)^3 \int\limits_{0}^{\infty} dp
p^2 \, \varphi_d(p) \int\limits_{0}^{\infty} dp' p'^2 \, \varphi_d(p')
\frac{1}{p_{>}} \frac{1}{4\pi},
\label{eq:2.37}
\end{equation}
where $p_{>}=max(p,p')$.

We now return to Eq.~(3.19), set $\tau=t-t^{\dagger}$, use the
relation given in Eq.~(\ref{eq:2.17e}) to the total NN cross section
and obtain
\begin{equation}
\langle \Phi|P(t-t^{\dagger})P G_0 tP|\Phi \rangle \rightarrow - 2 \,
\sigma^{NN}_{tot} \langle {\textstyle\frac{3}{4}}{\bf
q_0}|t({\textstyle\frac{1}{m}}({\textstyle\frac{3}{4}} q_0)^2) |
{\textstyle\frac{3}{4}}{\bf q_0} \rangle_s \; \langle \varphi_d
|\frac{1}{r^2}|\varphi_d \rangle \frac{1}{4\pi}.  \label{eq:2.38}
\end{equation}
Finally subtracting the conjugate complex according to Eq.~(2.6) and
 using the optical theorem in the two-body subsystem,
\begin{equation}
\langle {\textstyle {3 \over 4}} {\bf q}_0 | t | {\textstyle {3 \over
4}} {\bf q}_0 \rangle_s - \langle {\textstyle {3 \over 4}} {\bf q}_0 |
t | {\textstyle {3 \over 4}} {\bf q}_0 \rangle_s^* = - 2i {3 q_0 \over
4m} \left ( {1 \over 2 \pi} \right)^3 \sigma_{tot}^{NN},
\label{eq:2.38a}
\end{equation}
leads to
\begin{eqnarray}
\langle \Phi|P(t-t^{\dagger})P G_0 tP|\Phi \rangle & - & \langle
\Phi|P(t-t^{\dagger})P G_0 tP|\Phi \rangle^{\star} \nonumber \\ & &
\rightarrow i \frac{3 q_0}{m} \frac{1}{(2\pi)^3} \left(
\sigma^{NN}_{tot} \right)^2 \langle \varphi_d |\frac{1}{r^2}|\varphi_d
\rangle \frac{1}{4\pi}.
\label{eq:2.39}
\end{eqnarray}
Because of $(\sigma^{NN}_{tot})^2$ this expression is
of order $O(t^4)$.

It remains now to discuss the last term of second order in $t$ in
Eq.~(2.6) in the high energy limit. Here we encounter
$G_0-G_0^{\star}= -2i\pi \delta(E-H_0)$ and obtain
\begin{eqnarray}
\lefteqn{\langle \Phi | P t^{\dagger} P(G_0-G_0^{\star}) tP|\Phi
\rangle } \nonumber \hspace{1cm} \\ &\rightarrow& -4i \pi \, \left
|\langle {\textstyle\frac{3}{4}}{\bf q_0}|t({\textstyle
\frac{1}{m}}({\textstyle\frac{3}{4}} q_0)^2) |
{\textstyle\frac{3}{4}}{\bf q_0} \rangle_s \right |^2 \int d^3q \int
d^3 q' \; \langle \varphi_d |{\bf q} \rangle \, \langle {\bf
q}'|\varphi_d \rangle \nonumber \\ & & \times \,\delta \left
(-|\epsilon_d| - {\textstyle\frac{1}{m}}(q^2 +q'^2 +{\bf q}\cdot{\bf
q'}) + {\textstyle\frac{3}{2m}} {\bf q_0}\cdot ({\bf q}+{\bf q'})
\right ).
\label{eq:2.40}
\end{eqnarray}
The integral term can simply be related to the expression $I$ from
Eq.~(3.20). Similarly to Appendix A we work in momentum space and
obtain
\begin{eqnarray}
I'&\equiv& \int d^3q \int d^3 q' \; \langle \varphi_d |{\bf q} \rangle \,
\langle {\bf q}'|\varphi_d \rangle \; \delta \left (-|\epsilon_d| -
{\textstyle \frac{1}{m}}(q^2 +q'^2 +{\bf q}\cdot{\bf q'}) +
{\textstyle \frac{3}{2m}} {\bf q_0}\cdot ({\bf q}+{\bf q'}) \right )
\nonumber \\ &=& \int d^2 q_{\perp} \int d^2 q'_{\perp} \int d q_z
\int d q_{z'} \; \varphi_d \left (\sqrt{q_{\perp}^2 +q_z^2} \right )
\, \varphi_d\left (\sqrt{{q'}_{\perp}^2 +{q'}_z^2} \right )
\frac{1}{4\pi} \nonumber \\ & & \times \, \delta \left (-|\epsilon_d|
- {\textstyle \frac{1}{m}} (q_{\perp}^2 +{q'}_{\perp}^2 +{\bf
q}_{\perp}\cdot {\bf q'}_{\perp} +q_z^2 +{q'}_z^2) + {\textstyle
\frac{3}{2m}} q_0(q_z+q'_z) \right ). \label{eq:2.41}
\end{eqnarray}
Then using Eq.~(3.29) for the argument of the $\delta$-function we arrive
at the dominant term
\begin{eqnarray}
I' &\approx& m \int d^2 q_{\perp} \int d^2 q'_{\perp} \int ds' \int ds
\, \frac{4}{3} \frac{1}{2q_0} \delta(s) \, \varphi_d \left (
\sqrt{{q}_{\perp}^2 + (s'+{\textstyle \frac{1}{2}}s)^2} \right) 
\varphi_d \left ( \sqrt{{q'}_{\perp}^2 +(s'-{\textstyle
\frac{1}{2}}s)^2} \right) \frac{1}{4\pi} \nonumber \\ &=&
\frac{2m}{3q_0} \int d^2 q_{\perp} \int d^2 q'_{\perp} \int ds' \;
\varphi_d\left (\sqrt{q_{\perp}^2 + s'^2} \right )\, \varphi_d \left
(\sqrt{{q'}_{\perp}^2 + s'^2} \right ) \frac{1}{4\pi} \nonumber \\ &=&
- \frac{1}{i\pi} \, I.
\label{eq:2.42}
\end{eqnarray}
The last step follows from Eq.~(A.1), if one evaluates this integral
in a different way.  Thus we end up with
\begin{equation}
\langle \Phi|Pt^{\dagger}P(G_0 -G_0^{\star})tP| \Phi \rangle
\rightarrow -2i \frac{4m}{3q_0} (2\pi)^3 \, \left |\langle
{\textstyle\frac{3}{4}}{\bf q_0}
|t({\textstyle\frac{1}{m}}({\textstyle\frac{3}{4}} q_0)^2)
|{\textstyle\frac{3}{4}}{\bf q_0} \rangle_s \right |^2 \, \langle
\varphi_d|\frac{1}{r^2}| \varphi_d \rangle
\frac{1}{4\pi}. \label{eq:2.43}
\end{equation} 
Adding the above expression to Eq.~(\ref{eq:2.39}) and using Eq~(2.4) gives
 the asymptotic contribution of the terms second order in $t$ 
of Eq.~(2.6) to the total n+d cross section as 
\begin{eqnarray}
\sigma^{nd}_{tot}|_{2nd\:{\rm order}}& =&- 2 \left(
\sigma_{tot}^{NN}\right)^2 \langle \varphi_d|\frac{1}{r^2}|\varphi_d
\rangle \frac{1}{4\pi}\\ \nonumber & & + \, (2\pi)^6
\left(\frac{4m}{3q_0}\right)^2 \left |\langle {\textstyle \frac{3}{4}}{\bf
q_0} |t ({\textstyle \frac{1}{m}}({\textstyle \frac{3}{4}} q_0)^2) |
{\textstyle \frac{3}{4}}{\bf q_0} \rangle_s \right |^2 \, \langle
\varphi_d|\frac{1}{r^2}| \varphi_d \rangle
\frac{1}{4\pi}. \label{eq:2.44}
\end{eqnarray}
If we use the optical theorem in the 2-body system for the imaginary
 part occurring in $|\langle {\textstyle \frac{3}{4}}{\bf q_0}
 |t({\textstyle \frac{1}{m}}({\textstyle \frac{3}{4}} q_0)^2)
 |{\textstyle \frac{3}{4}}{\bf q_0} \rangle_s|^2$, we obtain
\begin{eqnarray}
\sigma^{nd}_{tot}|_{2nd\:{\rm order}} &\rightarrow& - \left(
\sigma_{tot}^{NN}\right)^2 \langle \varphi_d|\frac{1}{r^2}|\varphi_d
\rangle \frac{1}{4\pi}  \nonumber \\ & & + \, (2\pi)^6
\left(\frac{4m}{3q_0}\right)^2 \left( Re \langle
{\textstyle \frac{3}{4}}{\bf q_0} |t
({\textstyle \frac{1}{m}}({\textstyle \frac{3}{4}} q_0)^2)
|{\textstyle \frac{3}{4}}{\bf q_0} \rangle_s \right)^2 \langle
\varphi_d|\frac{1}{r^2}|\varphi_d \rangle
\frac{1}{4\pi}. \label{eq:2.45}
\end{eqnarray}

In summary, we obtain in the high energy limit for the total n+d cross
section based on the terms of first and second order in $t$ in the
expression for $ Im \langle \Phi |U| \Phi \rangle$
\begin{eqnarray}
\sigma^{nd}_{tot} &=& \sigma^{nd}_{tot}|_{1st\:{\rm order}} +
 \sigma^{nd}_{tot}|_{2nd\:{\rm order}} \nonumber \\ &=& 
2 \, \sigma_{tot}^{NN} \nonumber \\
 & & + \, (2\pi)^6 \left(\frac{4m}{3q_0}\right)^2 \left( Re \langle
 {\textstyle \frac{3}{4}}{\bf q_0}
 |t({\textstyle \frac{1}{m}}({\textstyle \frac{3}{4}} q_0)^2)
 |{\textstyle \frac{3}{4}}{\bf q_0} \rangle_s \right)^2 \langle
 \varphi_d|\frac{1}{r^2}|\varphi_d \rangle \frac{1}{4\pi}  \nonumber \\
 & & - \left( \sigma_{tot}^{NN}\right)^2 \langle
 \varphi_d|\frac{1}{r^2}|\varphi_d \rangle \frac{1}{4\pi} \nonumber \\
&=& 2 \, \sigma_{tot}^{NN} + O(t^2) - O(t^4). \label{eq:2.46}
\end{eqnarray}
As mentioned earlier, the first term, obtained from the first order
term in the two-nucleon $t$-matrix within the multiple scattering
expansion gives twice the NN total cross section, thus considering the
two nucleons in the deuteron as being free particles. The terms of
second order in $t$ within the multiple scattering expansion of
Eq.~(\ref{eq:2.6}) give rise to two correction terms.  The positive
term in Eq.~(3.43), being of $O(t^2)$, enhances the contribution of
the first term.  It is tempting to consider this term as an
anti-shadowing effect.  The third, negative term in Eq.~(3.43), which
is due to $(\sigma_{tot}^{NN})^2$ of $O(t^4)$, reduces the value of
the two total NN cross sections and its action is naturally called a
shadowing effect.  In the Glauber approximation
\cite{glauber,joachain,henley,abers} one arrives formally at the same
result. However, it should be pointed out that Eq.~(3.43) contains the
symmetrized two-body $t$-matrix elements and thus implicitly forward
and backward amplitudes.  We think that our derivation based on the
multiple scattering expansion of the Faddeev formulation together with
the momentum space treatment is a viable alternative.  It uses
standard and transparent evaluations of integrals, which could also be
calculated numerically without going to the high energy limit.

\section{Application}
In this Section we want to give a numerical illustration of the
results derived in the previous section.  We use two NN model forces
of Malfliet-Tjon type \cite{mft}, one being purely attractive and the
other having attractive and repulsive parts.  Both are of Yukawa type
and given as
\begin{equation}
V({\bf q'},{\bf q})= \frac{1}{2\pi^2}\left( \frac{V_R}{({\bf q'}- {\bf
      q})^2 + \mu_R^2} - \frac{V_A}{({\bf q'}-{\bf q})^2 + \mu_A^2}
      \right).  \label{eq:4.0}
\end{equation}
The parameters are given in Table~I. As we saw in the last section,
there are two momenta which control the asymptotic expansion. The
first is the initial projectile momentum ${\bf q_0}$, which appears as
$\frac{3}{4}{\bf q_0}$ in the three-body context. The second is the
`typical' deuteron momentum. For the two model potentials we display
in Fig.~1 the corresponding deuteron wave functions. We see that in
both cases the wave functions drop by about a factor of 10 within
0.6~fm$^{-1}$.  Thus we expect that for $\frac{3}{4} q_0$ sufficiently
larger than this value the asymptotic expressions derived in the
previous section should be valid.  In order to get quantitative
insight into the onset of the validity of the asymptotic expressions
one should evaluate the terms in the multiple scattering series
exactly. Strictly spoken this amounts to solving the Faddeev equation,
Eq.~(2.2). As a first step we exactly evaluate here the term in
Eq.~(2.10), which is in first order in $t$. Using Eq.~(3.6) one finds
the following exact form
\begin{equation}
\langle \Phi|P(t-t^{\dagger})P|\Phi\rangle = \frac{8i}{\pi}
\int d^3p \; \varphi_d^2(2|{\bf p}-{\textstyle \frac{3}{4}}{\bf q_0}|) \,
 Im \langle {\bf p} |t(\varepsilon)|{\bf p}\rangle_s \label{eq:4.2}
\end{equation}
where
\begin{equation}
\varepsilon = \epsilon_d + {\textstyle \frac{1}{m}} p^2 -
 {\textstyle \frac{4}{m}} ( {\bf p} - {\textstyle \frac{3}{4}} {\bf
 q_0})^2 \label{eq:4.3}
\end{equation}

Introducing as variables explicitly the magnitudes of the momentum
vectors ${\bf p}$ and ${\bf q_0}$ and the angle between them, $x={\hat
{\bf p}} \cdot {\hat {\bf q}}{\bf {_0}}$, we obtain for the n+d total cross
section in the first order in $t$
\begin{equation}
\sigma^{nd}_{tot}|_{1st\:{\rm order}} = -(2\pi)^3 \frac{4 m}{3 q_0} \, 8
\int\limits_{-1}^1 dx \int\limits_0^{\infty} dp \; p^2 \; \varphi_d^2
\left ( 2 \sqrt{p^2 +({\textstyle \frac{3}{4}}q_0)^2 -
{\textstyle \frac{3}{2}}pq_0 x} \right )\; Im \:
t_s(p,p,\varepsilon). \label{eq:4.4}
\end{equation}

\noindent
The integral over the imaginary part of the forward scattering
amplitude receives contributions from the region $\varepsilon > 0$ and
the deuteron pole.  The condition $\varepsilon > 0$ limits the
integration regions in $x$ as well as $p$. The $x$-integration has to
be carried out only between $x_{min}=\sqrt{\frac{3}{4}
+\frac{1}{3}\frac{m|\epsilon_d|} {q_0^2}}$ and $x_{max}=1$, and
$p$-values given by
\begin{equation}
p_{max,min}= q_0 x \pm \sqrt{q_0^2 x^2 - {\textstyle \frac{3}{4}}q_0^2 -
  {\textstyle \frac{1}{3}} m|\epsilon_d|}. \label{eq:4.5}
\end{equation}
The pole contribution is obtained from an integration over $x$ with
similar boundaries at fixed $p$-values.

The off-shell NN scattering amplitude is determined by firstly solving
the two-body Lippmann-Schwinger equation exactly in three dimensions,
i.e. without partial wave expansion \cite{3dt}.  The quantities
entering the integral in Eq.~(\ref{eq:4.4}) are off-shell $t$-matrix
elements for equal magnitudes of the momenta. Thus the integration
requires a two-dimensional interpolation of the imaginary part of the
forward scattering amplitude in the variables $p$ and
$\varepsilon$. We use standard b-splines for the numerical calculation
\cite{deboer}.

The first order term of the total n+d cross sections for the two
potential models MT-III and MT-IV is shown in Figs.~2 and 3 as
function of $\frac{3}{4} q_0$
and compared to its asymptotic limit $2
\sigma_{tot}^{NN}$ given in Eq.~(3.15).  A closer inspection reveals
that this limit is reached at nucleon laboratory energies of about 300
(600) MeV within 1\% for MT-III (MT-IV).  This energy is fairly high
so that the simple potential picture will no longer be valid and
relativistic features will be important, including meson production.
These absorption processes will of course change our results as is
known from studies within the Glauber formalism
\cite{joachain,henley}.  Nevertheless we think it should be of
theoretical interest to see the results within a pure potential
picture. We leave the exact solution of the Faddeev equation,
Eq.~(2.2), to a future study and estimate now the resulting
rescattering terms via the asymptotic expressions contained in
Eq.~(3.43). They are expected to be a good approximation around
300~MeV and above for the potential containing the realistic feature
of repulsion.

The deuteron matrix elements for the two potentials entering in the
asymptotic expression of Eqs.~(3.32) and (\ref{eq:2.46}) are given in
Table~I.

We start our investigation with the MT-III potential, which has a
short range repulsion and a intermediate range attraction, and thus
contains some realistic features of the NN force.  In Fig.~4 the
three different terms of the right hand side of
Eq.~(\ref{eq:2.46}) are shown separately as function of the nucleon
laboratory energy together with their partial sums building up the n+d
total cross section in the high energy limit. We see that the
rescattering term of $O(t^2)$ is quite
large and can not be neglected in relation to the leading term
2$\sigma_{tot}^{NN}$ in the whole energy range shown.  For a more
quantitative inspection the three different terms are listed
separately in Table~II for the higher energies.  The first column
gives twice the NN total cross section, the second column shows the positive
term, which is also of $O(t^2)$ like the total NN cross section and
the third column gives the negative term, which is of $O(t^4)$, and
which causes shadowing.  We see from the table that the positive term
of $O(t^2)$, which results from the second order in $t$ of the
multiple scattering expansion of the elastic forward scattering
amplitude $\langle \Phi|U|\Phi \rangle$ is relatively large.
Already at 100 MeV the shadowing term is much
smaller than the other two terms and drops of course quickly due to
its energy dependence $O(\frac{1}{q_0^4})$. From these numbers we see
that the rescattering corrections are quite large in this pure
potential picture. This is even more pronounced for the purely
attractive potential MT-IV as shown in Fig.~5.  The corresponding
terms are displayed in Table III. We see that the positive term of
$O(t^2)$, which results from the second order in $t$ of the multiple
scattering expansion of the elastic forward scattering amplitude
$\langle \Phi|U|\Phi \rangle$ is quite large and more important than
the sum of the two NN total cross sections.

In the purely attractive potential model MT-IV
we are able to rigorously provide the extreme
high energy limit of the n+d total cross section.  In this limit the
Born term for the NN $t$-matrix gives the only contribution and we can
replace the $t$-matrix element in Eq.~(3.41) 
with the potential matrix element, which
is real,
\begin{equation}
\langle {\textstyle \frac{3}{4}}{\bf q_0} |t
({\textstyle \frac{1}{m}}({\textstyle \frac{3}{4}}q_0)^2)
|{\textstyle \frac{3}{4}}{\bf q_0} \rangle_s \rightarrow \langle
{\textstyle \frac{3}{4}}{\bf q_0} |V|{\textstyle \frac{3}{4}}{\bf q_0}
\rangle_s.
\label{eq:4.1}
\end{equation}
This expression can be
related to the total cross section, and one finds after a simple
integration
\begin{equation}
\int d{\hat p} \, |\langle {\textstyle\frac{3}{4}}q_0{\hat p}
|V|{\textstyle\frac{3}{4}}{\bf q_0} \rangle_s|^2 \rightarrow 4\pi
\frac{8}{9q_0^2} \, \mu_A^2 \, \langle
{\textstyle\frac{3}{4}}{\bf q_0} |V|{\textstyle\frac{3}{4}}{\bf
q_0}\rangle^2_s.
\label{eq:4.6}
\end{equation}
Using the definition of the NN total cross section we obtain
\begin{equation}
\langle {\textstyle \frac{3}{4}}{\bf q_0}
|V| {\textstyle \frac{3}{4}} {\bf q_0} \rangle^2_s \rightarrow \left(
\frac{3q_0}{m} \right)^2 \frac{1}{(2\pi)^5
 \frac{1}{2\mu_A^2}} \,\sigma_{tot}^{NN}. \label{eq:4.7}
\end{equation}
As a consequence one can write  Eq.~(\ref{eq:2.46}) as
\begin{equation}
\sigma^{nd}_{tot} \rightarrow 2 \sigma_{tot}^{NN} \left[ 1 + 2 \langle
\varphi_d|\frac{1}{\mu^2_A r^2}|\varphi_d \rangle \right] - \left(
\sigma_{tot}^{NN}\right)^2 \langle \varphi_d|\frac{1}{r^2}| \varphi_d
\rangle \frac{1}{4\pi}.  \label{eq:4.8}
\end{equation}
For the potential MT-IV the term additive to 1 is 3.166 (using the
quantities given in Table~I) and therefore the total NN cross section
is multiplied by 8.33 if the exact limit is reached. This number is
substantially larger than the naive expectation, which would be twice
the NN cross section. Numerically this asymptotic limit is reached
around 4000 MeV nucleon laboratory energy within 2\% as we see from
the last column in Table~III. There we display the ratio  of the
high energy limit obtained from Eq.~(\ref{eq:2.46}) and the exact
limit from Eq.~(\ref{eq:4.8})
\begin{equation}
R \equiv 
\frac{2 \, \sigma_{tot}^{NN}+(2\pi)^6 \left(\frac{4m}{3q_0}\right)^2
\left( Re \langle
\frac{3}{4}{\bf q_0} |t(\frac{1}{m}(\frac{3}{4} q_0)^2)
|\frac{3}{4}{\bf q_0} \rangle_s \right)^2
\langle \varphi_d|\frac{1}{r^2}|\varphi_d \rangle \frac{1}{4\pi}}
{2 \, \sigma_{tot}^{NN}\left[1 + 2 \langle \varphi_d|\frac{1}{\mu^2_A
r^2}|\varphi_d \rangle \right] }, \label{eq:4.9}
\end{equation}
where contributions of $O(t^4)$ to the total cross section are
neglected. The factor $F_{\rm MT-IV}$ shown in Table~III is explained below.

For the MT-III potential there is interference among the repulsive and
attractive parts of the potential and the connection between the
potential matrix element in forward direction and the total NN cross
section is more involved.  Since $\sigma_{tot}^{NN} =
O(\frac{1}{q_0^2})$ and the real part of the forward scattering
amplitude $t$ approaches a constant, both positive terms in Eq.~(3.43)
can again be combined as $\sigma_{tot}^{NN} \times F$ and we
determined the factor multiplying $\sigma_{tot}^{NN}$ as
$F_{\rm MT-III}=2.83$, as shown in the last column of Table~II. Because of
the repulsion contained in the model MT-III the asymptotic limit is
reached much later compared to the purely attractive potential model 
MT-IV.

In both cases the rigorous asymptotic limits are
\begin{equation}
\sigma^{nd}_{tot} \rightarrow \sigma^{NN}_{tot} \times F,
\end{equation}
where $F$ is either $F_{\rm MT-III}$ or $F_{\rm MT-IV}$.
The factors F contain the deuteron matrix element $\langle
\varphi_d|\frac{1}{r^2}|\varphi_d \rangle$ and numerical as well as
potential parameter constants. 
The factors $F$ are larger
than 2 for both potential models,
and thus the rescattering process enhances the asymptotic cross
section over the sum of the two NN cross sections, the shadowing has
vanished already  before that.

We started from an expansion of the elastic n+d forward scattering
amplitude $\langle \Phi|U|\Phi \rangle $ in powers of the NN
$t$-matrix. Due to the optical theorem the term first order in $t$ ended
up as a term second order in $t$ in the total n+d cross section. 
The terms second order in $t$ in the elastic forward scattering
amplitude provide two terms in the total n+d cross section, one of
second order and one of fourth order in $t$.  The one second order in
$t$ behaves like $O(\frac{1}{q_0^2})$, whereas the one of fourth order
in $t$ decreases like $O(\frac{1}{q_0^4})$ in the limit $q_0$ going to
infinity. It is therefore natural to group the terms together according
to their power in $t$ as we did in Eq.~(\ref{eq:2.46}), which coincides with
their energy dependence. Thus the shadowing effect will disappear
faster than the anti-shadowing effect.  For the two model forces
considered these anti-shadowing effects modify the naive expectation
that the total cross section for n+d scattering tends to twice the
total NN cross section. The true asymptotic result is larger than
twice the NN cross section, which means that the two nucleons in the
deuteron can never be considered to be independent.  Finally it is
obvious that the terms in Eq.~(2.6), which are third order in $t$ can
not cancel the terms $O(t^2)$ in the total n+d cross section. They may
however modify the shadowing effect, since there will be also
contributions of order $O(t^4)$ to the total n+d cross section. Thus
the asymptotic value of the n+d total cross section in our model of
spinless nucleons interacting by local forces is the first term on the
right hand side of Eq.~(4.9) in case of the purely attractive Yukawa
potential MT-IV, and $\sigma^{NN}_{tot} \times F$ with $F=2.83$ for the
potential model MT-III. These are exact results for our potential
models.

\section{Summary}

In view of new precise measurements of the total n+d cross section at
energies above 100 MeV nucleon laboratory energy and of precise
solutions of the Faddeev equations using modern NN forces
\cite{ndtotl} it is of interest to understand the leading rescattering
effects which modify the naive expectation that the total n+d
scattering cross section is the sum of the np and nn total cross
sections. In a model of spinless nucleons we investigate the first few
terms of the multiple scattering series for the forward elastic n+d
scattering amplitude resulting from the Faddeev equations in the high
energy limit. Although the treatment is purely nonrelativistic and
enters far into the region, where relativity is important, we think it
is interesting to know the asymptotic behavior at high energies for
pure potential models without absorption. Absorption processes
(particle productions) occurring in a relativistic context will
however change the results presented here \cite{glauber,henley}.  In
accordance to the naive expectation we extract from the second order
term in the multiple scattering series a shadowing effect proportional
to the total NN cross section squared, which is negative and reduces
the total n+d cross section.  However, we also find a positive term
proportional to the square of the real part of the NN forward
scattering amplitude, which decreases in energy only proportional to
the total NN cross section, whereas the shadowing term vanishes
faster. Both terms are proportional to the expectation value of
1/(nucleon distance)$^2$ with respect to the deuteron wave function.

In a numerical illustration using NN force models of Yukawa type the
positive term has as limit the total NN cross section multiplied by a
number larger than 2 (which would be the naive expectation).  For the
purely attractive potential this factor is 8.15, whereas for the
potential with the additional repulsion this factor is 2.83.  These
are exact results for our choice of potential models.  For both
potentials the negative shadowing effect decreases faster as function
of energy than the positive terms.

We expect that the asymptotic expressions given in Eq.~(3.43) start to
be valid around 300-600~MeV depending on the potential model employed.
Strictly speaking this can only be assured, if the Faddeev equation,
Eq.~(2.2), is solved exactly.  The estimate given in this work is
based on the exact evaluation of the term first order in the NN
$t$-matrix in the n+d elastic forward scattering amplitude.  From this
calculation we found the corresponding asymptotic limit starting to be
valid at the quoted energies.

The formal expression, Eq.~(3.43), has also been found in the context
of the Glauber approximation \cite{glauber,abers}. A simple
geometrical explanation of the deuteron matrix element has been given
in Ref.~\cite{glauber}.  Our derivation is different and follows
simply from the elastic forward n+d scattering amplitude evaluated in
leading orders in the NN $t$-matrix and taking the high energy limit
analytically.  A similar investigation including spin and isospin
degrees of freedom is in preparation.

\vfill

\acknowledgements This work was performed in part under the
auspices of the U.~S.  Department of Energy under contract
No. DE-FG02-93ER40756 with Ohio University, a grant of the Ohio Board
of Regents Research Challenge Program, the NATO Collaborative
Research Grant 960892, and the Research Contract 41324878(COSY-44) with
the Forschungszentrum J\"ulich. Two of the authors (Ch.E. and W.G.)
would like to thank the Institute for Nuclear Theory at the University
of Washington for their hospitality during some part of the work.  We
thank the Ohio Supercomputer Center (OSC) for the use of their
facilities under Grant No.~PHS206.


\appendix
\section{Derivation of the momentum space form of the integral I}

In this Appendix we derive the limiting momentum space form of the
integral I given in Eq.~(3.20). Using the leading expression for the
denominator as given in Eq.~(3.29) the expression in Eq.~(3.20) can be
rewritten as
\begin{eqnarray}
I &\rightarrow& \frac{2m}{3q_0} \int d^3q \int d^3q' \; \frac{
\varphi_d(q) \, \varphi_d(q')}{q_z+q'_z +i\varepsilon} {1 \over 4 \pi}
\nonumber \\ &=& \frac{2m}{3q_0} (2\pi)^2 \int\limits_0^{\infty}dq 
q^2\int\limits_0^{\infty}dq'  q'^2 \int\limits_{-1}^1 dx
\int\limits_{-1}^1 dx' \,\frac{\varphi_d(q) \,\varphi_d(q')} {qx
+q'x'+i\varepsilon} {1 \over 4 \pi}.
\end{eqnarray}
The x and x' integrals are elementary and one ends up directly with
the expression given in Eq.~(3.33).




\begin{table}
\caption{ Parameters of the Malfliet-Tjon type potentials.
As conversion factor we use units such that $\hbar c$=197.3286 MeVfm=1.}
\begin{tabular}{c|cccc|c}
\mbox{  } & $V_A$ & $\mu_A$ [MeV] & $V_R$  & $\mu_R$ [MeV] &
$\langle \varphi_d|\frac{1}{r^2}|\varphi_d \rangle$ [fm$^{-2}$] \\
\hline
 MT-IV  & 0.3303  & 124.91  &   -    &  -      & 0.6343 \\
 MT-III  & 3.1769  & 305.86  & 7.291  & 613.69  & 0.3090 \\
\end{tabular}
\end{table} 

\begin{table}
\caption{\label{tabmtiii}Contributions of the different leading order terms of the
MT-III potential to the total n+d cross section as function of the NN
laboratory energy.  The last column shows the factor $F_{\rm MT-III}$,
which multiplies $\sigma_{tot}^{NN}$ and approaches a constant for
higher energies.}
\begin{tabular}{ccccc}
$E_{lab}$ [MeV] & 2 $\sigma_{tot}^{NN}$ [fm$^2$] &  O($t^2$) [fm$^2$] &
   O($t^4$) [fm$^2$] & $F_{\rm MT-III}=\frac{ 2 \sigma_{tot}^{NN} + O(t^2)}
   {\sigma_{tot}^{NN}}$ \\ \hline
  50  &  63.945 &  22.845  &  -25.218  &  2.71 \\
  100 &  23.438 &  19.410  &  -3.388   &  3.66 \\
  300 &  9.885  &  8.376   &  -0.603   &  3.69 \\
  500 &  7.314  &  5.027   &  -0.330   &  3.37 \\
 1000 &  4.577  &  2.595   &  -0.129   &  3.13 \\
 1500 &  3.400  &  1.769   &  -0.071   &  3.04 \\
 2000 &  2.725  &  1.341   &  -0.046   &  2.98 \\
 2500 &  2.283  &  1.078   &  -0.032   &  2.94 \\
 3000 &  1.970  &  0.899   &  -0.024   &  2.91 \\
 3500 &  1.738  &  0.771   &  -0.019   &  2.88 \\
 4000 &  1.557  &  0.673   &  -0.015   &  2.86 \\
 4500 &  1.414  &  0.597   &  -0.012   &  2.84 \\
 5000 &  1.296  &  0.536   &  -0.010   &  2.83 \\
\end{tabular}
\end{table}

\begin{table}
\caption{Contributions of the different leading order terms of the
MT-IV potential to the total n+d cross section as function of the NN
laboratory energy.  The explicit expression for $R$ is given in
Eq.~(\ref{eq:4.9}). The definition for $F_{\rm MT-IV}$ is the same
as in Table \ref{tabmtiii}.}
\begin{tabular}{cccccc}
$E_{lab}$ [MeV] & 2 $\sigma_{tot}^{NN}$ [fm$^2$] &  O($t^2$) [fm$^2$] &
   O($t^4$) [fm$^2$] & $F_{\rm MT-IV}$ & $R$ \\ \hline
  50 & 82.057  & 101.812  &  -84.392 & 4.48 & 0.54  \\
 100 & 39.035  &  76.478  &  -19.098 & 5.92 & 0.71  \\
 300 & 12.076  &  31.616  &  -1.828  & 7.24 & 0.87  \\
 500 &  7.040  &  19.590  &  -0.621  & 7.57 & 0.91  \\
1000 &  3.411  &   9.989  &  -0.146  & 7.86 & 0.94  \\
1500 &  2.242  &   6.693  &  -0.063  & 7.97 & 0.96  \\
2000 &  1.669  &   5.035  &  -0.035  & 8.03 & 0.96  \\
2500 &  1.329  &   4.035  &  -0.022  & 8.07 & 0.97  \\
3000 &  1.097  &   3.349  &  -0.015  & 8.08 & 0.97  \\
3500 &  0.946  &   2.894  &  -0.011  & 8.12 & 0.97  \\
4000 &  0.817  &   2.513  &  -0.008  & 8.15 & 0.98  \\
4500 &  0.729  &   2.240  &  -0.007  & 8.15 & 0.98  \\
5000 &  0.661  &   2.034  &  -0.006  & 8.15 & 0.98  \\
\end{tabular}
\end{table}

\pagebreak
 
\noindent
\begin{figure}
\caption{The absolute values of the bound state wave functions 
$\varphi_d(k)$ 
as calculated from the two potential models MT-III and MT-IV given in
Table~I. \label{fig1}}
\end{figure}

\noindent
\begin{figure}
\caption{The contribution of the first order term of the multiple
scattering series for the n+d total cross section calculated exactly
from the MT-III potential model as function of the asymptotic momentum
$\frac{3}{4}q_0$ (solid line). The dashed line represents the
corresponding contribution of the first order term in the high energy
limit. The variable
$\frac{3}{4} q_0 = \sqrt{\frac{m}{2} E_{lab}}$ is chosen such that the
kinetic laboratory energies for the two and three body systems are the
same.
\label{fig2}}
\end{figure}

\noindent
\begin{figure}
\caption{Same as Fig.~2, except that the two-body potential employed
is the MT-IV potential. \label{fig3}}
\end{figure}

\noindent
\begin{figure}
\caption{Contributions to the n+d total cross sections as given by the
MT-III potential as function of the corresponding NN laboratory
energy.  The solid line shows twice the total NN cross
section. Successively added to this is the positive contribution
(antishadowing) of $O(t^2)$ (long dashed) and the negative
contribution (shadowing) of $O(t^4)$ (dash-dotted).  The magnitudes of
those contributions are shown separately as dotted line for the
positive and as short dashed line for the negative term. \label{fig4}}
\end{figure}

\begin{figure} 
\caption{Same as Fig.~4 except that the two-body potential employed
is the MT-IV potential. \label{fig5}}
\end{figure}

\newpage

\input{psfig}

\centerline{\hspace{35mm} \psfig{file=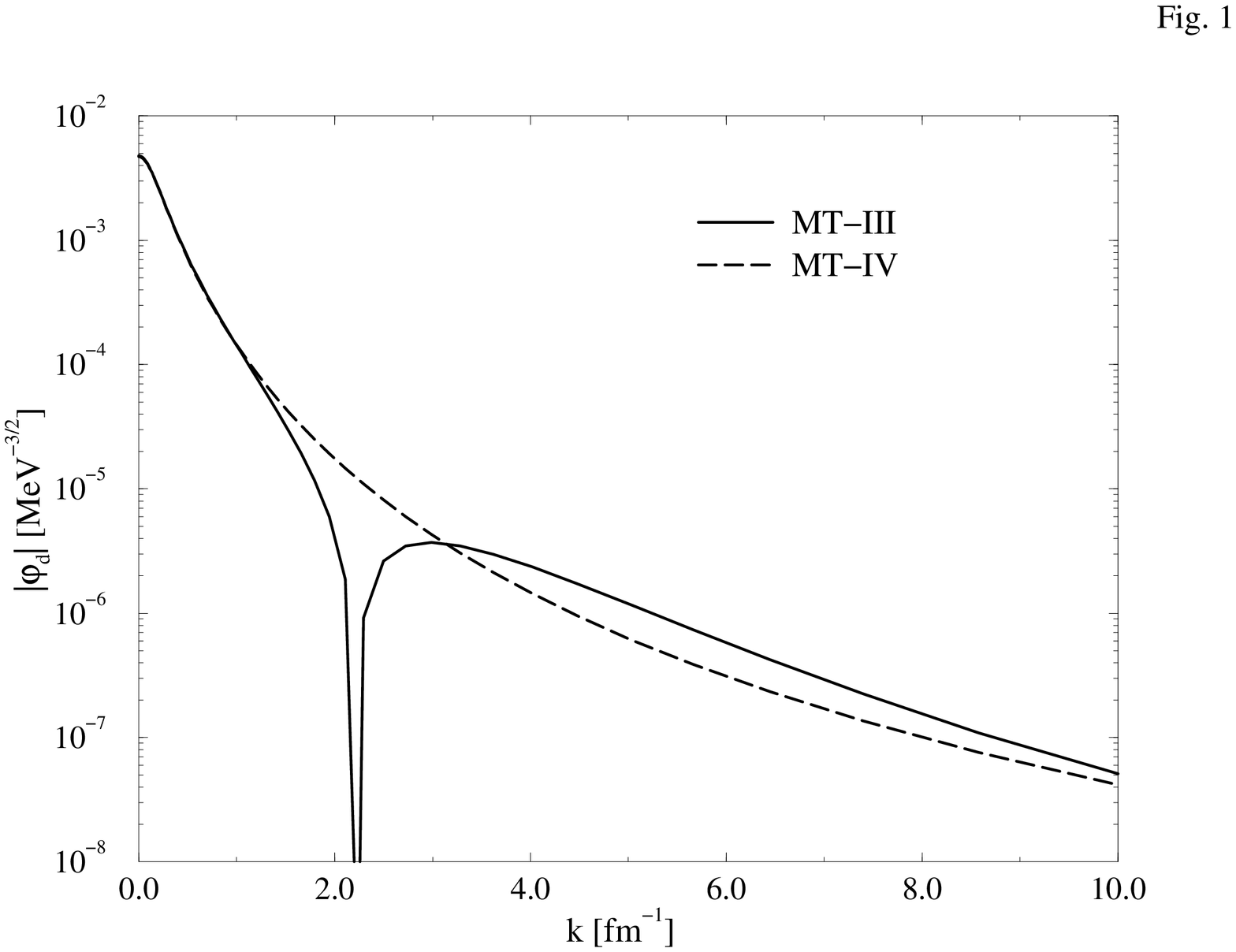,width=150mm,angle=-90}}
\centerline{\hspace{35mm} \psfig{file=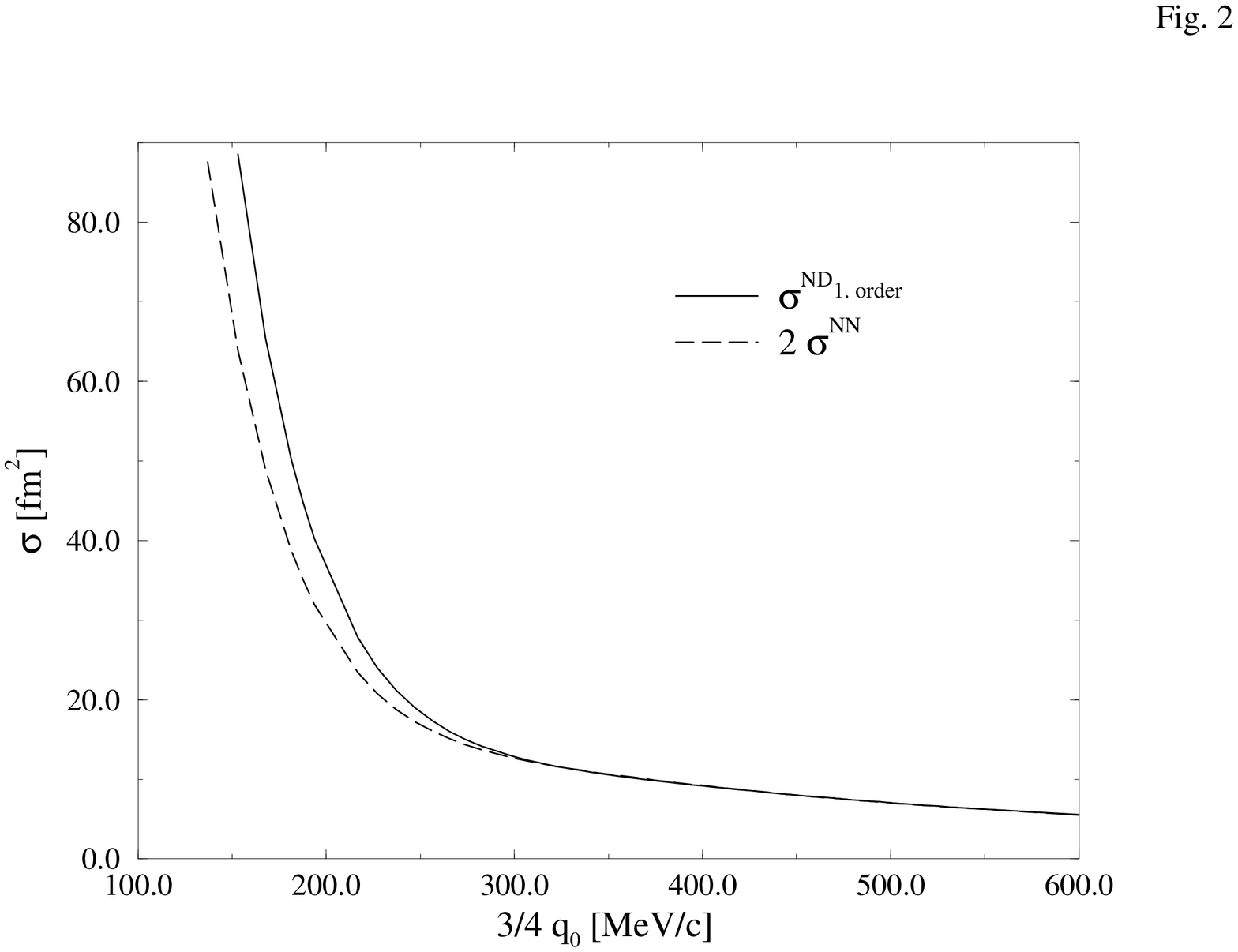,width=150mm,angle=-90}}
\centerline{\hspace{35mm} \psfig{file=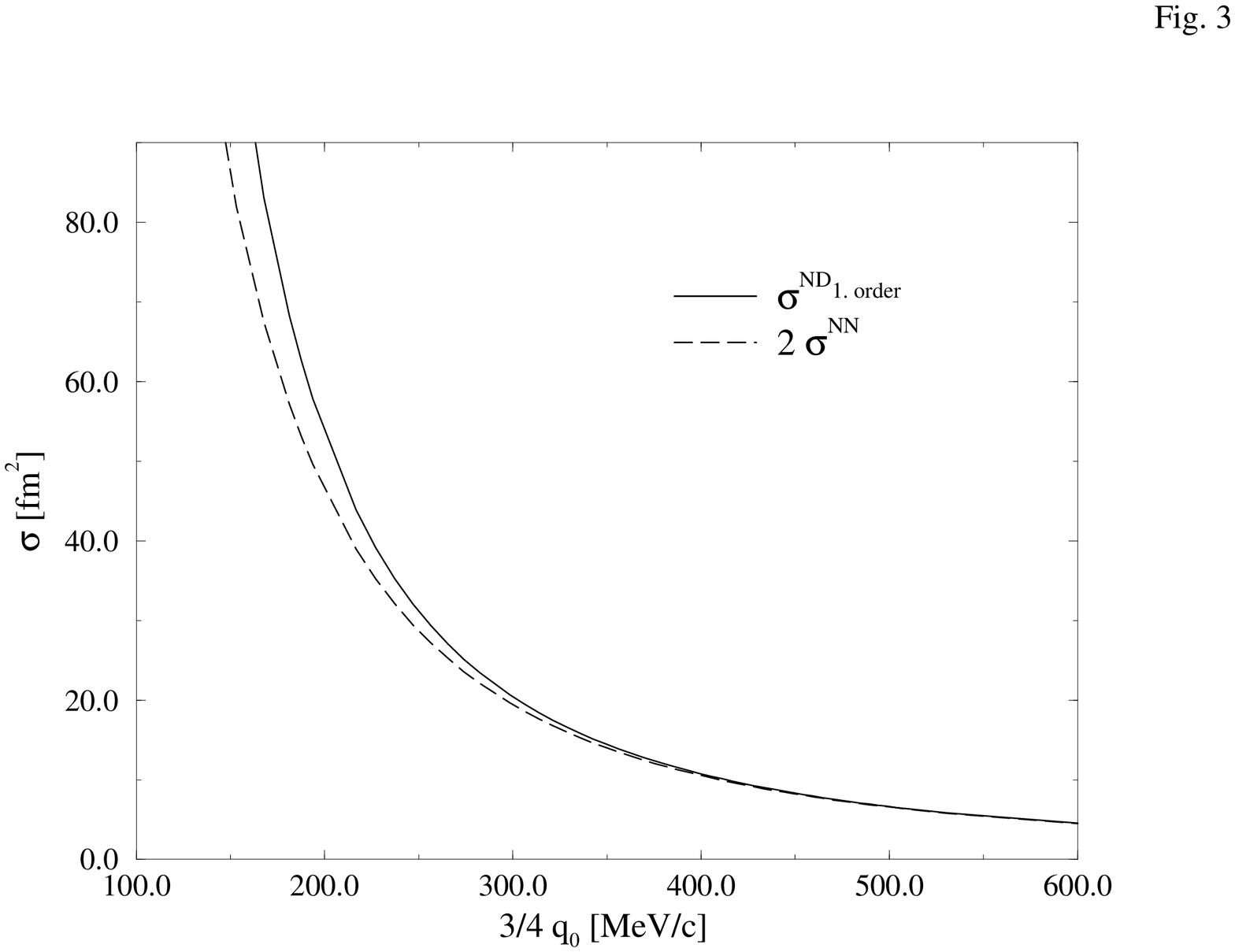,width=150mm,angle=-90}}
\centerline{\hspace{35mm} \psfig{file=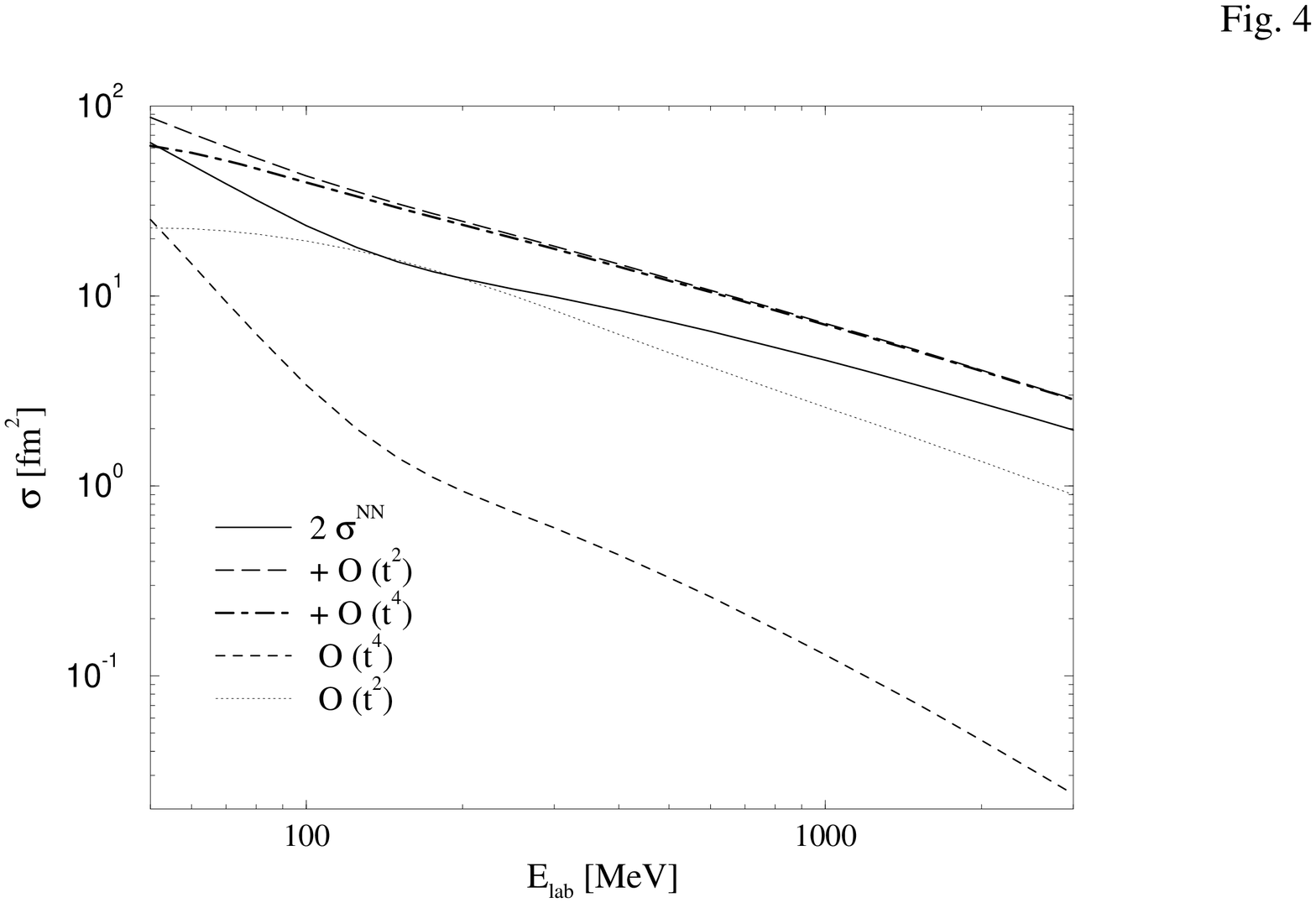,width=150mm,angle=-90}}
\centerline{\hspace{35mm} \psfig{file=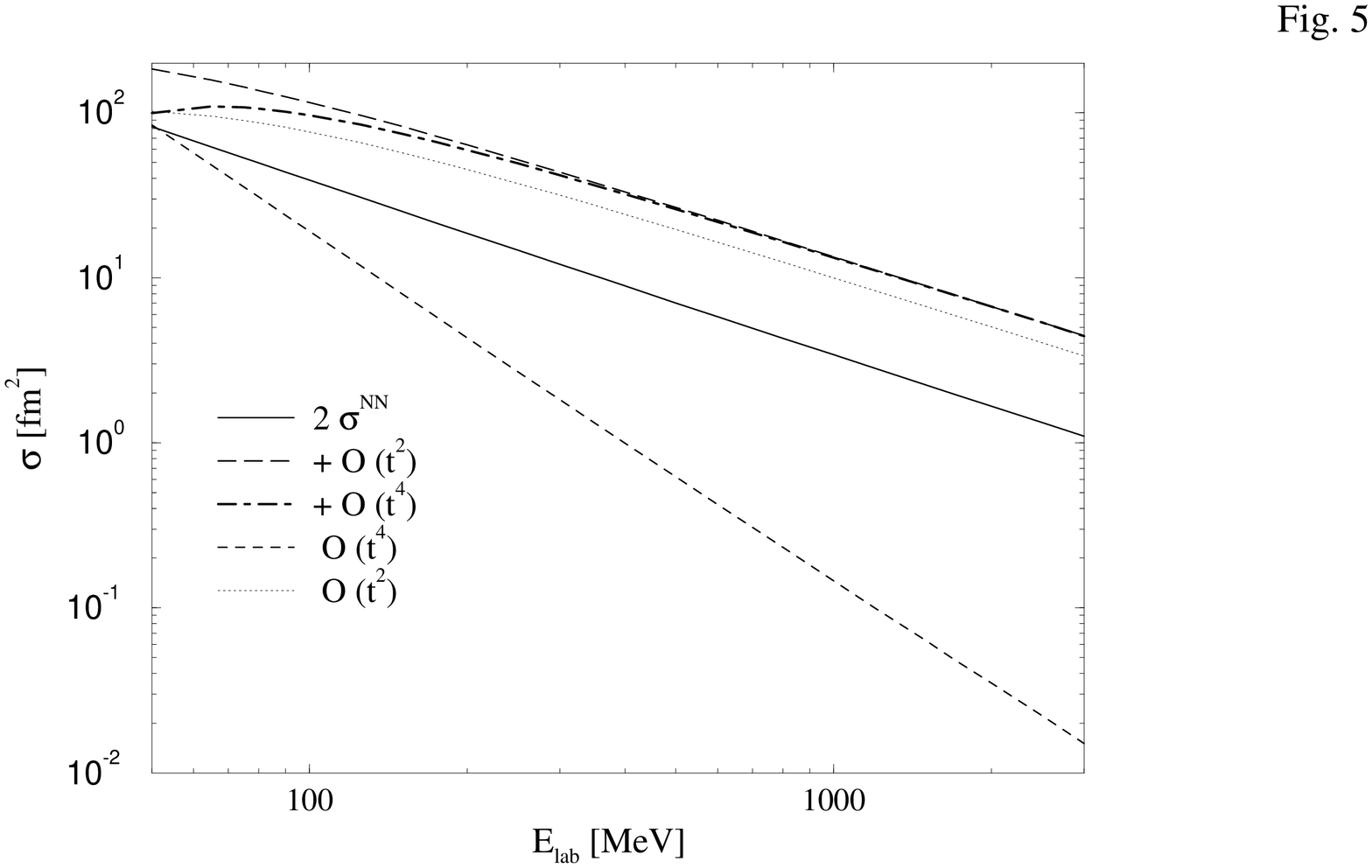,width=150mm,angle=-90}}

\end{document}